\documentclass[apj]{emulateapj}

\newcommand{\beq}{\begin{equation}}
\newcommand{\eeq}{\end{equation}}
\newcommand{\bey}{\begin{eqnarray}}
\newcommand{\eey}{\end{eqnarray}}

\newcommand{\pc}{\, {\rm pc} }
\newcommand{\kpc}{\, {\rm kpc} }
\newcommand{\msun}{M_\odot}
\newcommand{\Msun}{M_\odot} 
 \newcommand{\kms}{\, {\rm km \, s}^{-1} }

\newcommand{\grad}{{\bf \nabla}}

\begin{document}


\title{Loss of mass and stability of galaxies in MOND}

\author{Xufen Wu, HongSheng Zhao}\email{hz4@st-andrews.ac.uk}
\affil{SUPA, University of St Andrews, KY16 9SS, Fife, UK}
\author{Benoit Famaey}
\affil{Institut d'Astronomie et d'Astrophysique, Universit\'e Libre
de Bruxelles, Boulevard du Triomphe CP226, B-1050, Bruxelles, BELGIUM}
\author{G. Gentile}
\affil{University of New Mexico, Department of Physics and
Astronomy, 800 Yale Blvd NE, Albuquerque, New Mexico 87131, USA}
\author{O. Tiret, F. Combes}
\affil{Observatoire de Paris, LERMA, 61 Av. de l'Observatoire, 75014 Paris, France}
\author{G.W. Angus}
\affil{University of St Andrews, KY16 9SS, Fife, UK}
\author{A.C. Robin}
\affil{CNRS-UMR 6213, Institut UTINAM, Observatoire de Besan\c{c}on, BP1615, 
F25010 Besan\c{c}on cedex, France}


\begin{abstract}
The self-binding energy and stability of a galaxy in MOND-based
gravity are curiously decreasing functions of its center of mass
acceleration (of the order of $10^{-12}$ - $10^{-10}$m/s$^2$) towards neighbouring mass concentrations. A tentative indication of this breaking of the Strong Equivalence Principle in field galaxies is the RAVE-observed
escape speed in the Milky Way. Another consequence is that satellites of field galaxies will move on nearly Keplerian orbits at large radii (100 - 500 kpc), with a declining speed below the asymptotically constant naive MOND prediction. But conseqences of an environment-sensitive gravity are even more severe in clusters, where member galaxies accelerate fast: no more Dark-Halo-like potential is present to support galaxies, meaning that extended axisymmetric disks of gas and stars are likely unstable.  These predicted reappearance of asymptotic Keplerian velocity curves and disappearance of ``stereotypic galaxies" in clusters are falsifiable with targeted surveys. 
\end{abstract}

 \maketitle

\section{Introduction}

The exact nature of dark matter is an outstanding puzzle 
despite our ability to carry out increasingly realistic simulations. No astroparticle or gravitational theory so far can account for its various effects on both galactic and large scales satisfactorily (e.g., Zhao 2007).

In relatively isolated (field) galaxies, observations of a tight
correlation between the mass profiles of baryonic matter and dark
matter at all radii (McGaugh et. al. 2007; Famaey et. al. 2007a) are most often interpreted as supporting the modified
Newtonian dynamics (MOND; Milgrom 1983; Bekenstein \& Milgrom 1984). Indeed, without resorting to galactic dark matter,
this simple prescription reproduces (to amazing consistency) the
kinematics of galaxies over five decades in mass (e.g. Sanders \& McGaugh
2002; Bekenstein 2006; Milgrom \& Sanders 2003; Sanders \& Noordermeer 2007; Nipoti et al. 2007; Famaey, Bruneton \& Zhao 2007b; Gentile et. al 2007ab). The  recently devised relativistic counterparts of MOND (Bekenstein 2004; Zlosnik, Ferreira \&
Starkman 2007) also enabled to study the Cosmic Microwave Background (Skordis et al. 2006) as well as gravitational lensing of galaxies and galaxy clusters (e.g., Zhao et al. 2006; Chen \& Zhao 2006; Angus et al. 2007; Famaey et al. 2007c). However, it is also important to observationally and theoretically examine how
internal properties of galaxies like their rotation curve and morphology should change with the environment.

Indeed MOND-based theories generate different degrees of dark matter-like
effects depending on the absolute acceleration.  Most galaxies are,
like the Milky Way, in the field where they accelerate slowly with
respect to the Cosmic Microwave Background, typically at a rate of
$0.01$$a_0$ to $0.03$$a_0$ (Famaey et al. 2007b; Milgrom
2002; Angus \& McGaugh 2007). But in X-ray clusters, galaxies
accelerate much faster, from $0.3a_0$ to $3a_0$ (Angus et. al.
2007; Pointecouteau, Arnaud \& Pratt 2005).

This external gravitational field has wider and more subtle
implications for the internal system in MOND than in Newton-Einstein
gravity, for the very reason that MOND breaks the Strong Equivalence
Principle. In particular, it is well known that MOND potentials are
logarithmic for isolated distributions of finite mass, and
consequently infinitely deep, but that the internal potential
becomes ``polarised Keplerian" at large distances (Bekenstein \&
Milgrom 1984; Zhao \& Tian 2005) when an external field is applied.

In this contribution, we {\it numerically} solve the MOND Poisson
equation for systems embedded in several different environments,
ranging from the field to galaxy clusters, and show (i) that for the
Milky way (embedded in a weak gravitational field), the local escape
speed is numerically compatible with the observations as
analytically predicted in Famaey et al. (2007b), (ii) that
rotation curves of Milky Way-like galaxies would have a rapid Keplerian
fall-off when residing close to the center of clusters, while this fall-off in field galaxies would happen at $100$ - $500$ kpc, and (iii) that usual Low Surface Brightness disks should not exist in MONDian clusters.

\section{Binding energy of an accelerating Milky Way}

Galaxies free-fall, but with slowly-changing systematic
(center-of-mass) velocity ${\mathbf v}_{com}(t)$. Their present
non-zero systematic velocity is mainly the accumulation of the
acceleration by the gravity from neighbouring galaxies over a Hubble
time. Consider, as a first approximation, that a galaxy is
stationary in a non-inertial frame (in the Galilean sense), which
free-falls with a ``uniform" systematic acceleration $\dot{\mathbf
v}_{com}=g_{\rm ext} \hat{X} = cst$ due to an external linear
potential, say, $-g_{\rm ext} X$ along the $X$-direction, where the
over-dot means time-derivatives. Let $\dot{\mathbf v}_{\rm
int}=(\ddot{X},\ddot{Y},\ddot{Z})$ be the peculiar acceleration of a
star-like test particle in the coordinates relative to the center of
a non-evolving galaxy internal mass density $\rho(X,Y,Z)$, then \beq
\dot{\mathbf v}_{\rm int} = {\bf g} - \dot{\mathbf v}_{com}=-\grad
\Phi_{\rm int}(X,Y,Z), \eeq where ${\bf g}$ is the absolute
acceleration ${\bf g}$ satisfying the MOND Poisson's equation \beq
\label{Poisson} -\nabla.[\mu(x) {\bf g} ] = 4 \pi G \rho(X,Y,Z),
\qquad x \equiv {|{\bf g}| \over a_0}. \eeq One can define an
``effective potential" $\Phi_{\rm int}(X,Y,Z)$ (called ``internal"
potential) and an ``effective energy" $E_{\rm eff}={v^2_{\rm int}
\over 2}+\Phi_{\rm int}(X,Y,Z)$, where $E_{eff}$ is conserved along
the orbit of the test particle effectively moving in a force field
$-\grad \Phi_{\rm int}(X,Y,Z)$, which is curl-free and
time-independent because the absolute gravity ${\bf g}$ is
curl-free, center-of-mass acceleration $\dot{\mathbf v}_{com}$ is
assumed a constant, and the galaxy density $\rho(X,Y,Z)$ is assumed
time-independent.

Far away from the center of the free-falling system, we have
$|\dot{\bf v}_{\rm int}| \ll |\dot{\bf v}_{\rm com}|$, hence $\mu
\rightarrow \mu_m \equiv \mu(\dot{\bf v}_{com}/a_0)  ={\rm cst}$,
and the equation reads (Bekenstein \& Milgrom 1984; Milgrom1986; Zhao \& Tian 2005; Zhao
\& Famaey 2006): \beq \nabla^2 \Phi_{\rm int} + \Delta {\partial^2
\over
\partial X^2} \Phi_{\rm int} \rightarrow 4 \pi G \rho/\mu_m, \eeq
where $Y,Z$ denote the directions perpendicular to the external
field $X$-direction, and $\Delta=[{\rm dln}\mu/{\rm
dln}x]_{x=|\dot{v}_{com}|/a_0}$ is a dilation factor (note that $1
\le 1+\Delta \le 2$).  So at large radii  where the external field
dominates, and the equation is linearizable, the potential satisfies
a mildly anisotropic Poisson equation, and the solution at large
radii \footnote{throughout the paper, by ``large radii" we mean a
distance large enough to neglect the internal field $g_{\rm int} \ll
\dot{v}_{com} =g_{\rm ext}$ but small enough that the external field
$g_{\rm ext}$ can be treated as constant} goes to \beq \Phi_{\rm
int}^{\infty}(X,Y,Z) = -{GM_{\rm int} \over \mu_m
\sqrt{(1+\Delta)(Y^2+Z^2)+X^2 + s^2}}, \eeq where we included a
softening radius $s$, comparable of the half-light radius of a
galaxy.  Hence the internal potential $\Phi_{\rm int}$ is finite,
and approaches zero at large radii.

The escape speed at any location ${\bf r}$ in the system can then be meaningfully
defined by \beq
0 = E_{\rm eff}={v^2_{\rm esc}(X,Y,Z) \over 2}+\Phi_{\rm int}(X,Y,Z). \eeq
Such escape speed is a scalar independent of ``the path to escape"
because the ``effective energy" $E_{\rm eff}$ is conserved, and
a particle with $E_{\rm eff}$ equal zero (the maximum value of $\Phi_{\rm int}$)
will reach infinite distance from the system, and never return, hence will be lost
into the MOND potential of the background (from which it cannot escape).
However, equal escape speed contours
across a disk galaxy are generally not axisymmetric, meaning the escape speed on
opposite symmetric locations of the Galaxy differ.

Hereafter, we {\it numerically} solve Eq.~\ref{Poisson} using the
MOND Poisson solver developed by the Bologna group (Ciotti,
Londrillo, \& Nipoti 2006); the results based on spherical grids are
also confirmed with the cartesian grid-based code of the Paris group
(Tiret \& Combes 2007) with very different spatial resolutions. We
program in the mass density of the internal system, solving the MOND
Poisson equation as if it were isolated, except for requiring a
boundary condition on the total gravity as $ -{\mathbf g}
\rightarrow  g_{\rm ext} \hat{\mathbf X} - {\mathbf\grad} \Phi_{\rm
int}^{\infty}(X,Y,Z) $ on the last grid point $(X,Y,Z)$. Note
finally that in our models hereafter, we use the parametric
$\mu$-function $ \mu(x) = x/(1+x)$,
which fits well the rotation curve of the Milky Way (Famaey \&
Binney 2005), as well as external galaxies (Famaey et. al.
2007a; Sanders \& Noordermeer 2007).

We use the Besan\c{c}on Milky Way Model (Robin et. al 2003)
to simulate High Surface Brightness (HSB) galaxies. This model is a
realistic representation of the Galaxy, explaining currently
available observations of different types (photometry, astrometry,
spectroscopy) at different wavelengths. The stellar populations
included in the model are : a thin disk made of seven isothermal
layers each having a different age, between 0.1 and 10 Gyr; a
11-Gyr-old thick disk with a modified exponential density law, a
spheroid with a power law density, slightly flattened, a prolate old
bulge modeled by a triaxial density law. We logically take the dark
matter halo away from our simulations. We then apply the MOND
Poisson solver using $512 \times 64 \times 128$ grid points where
the grid points in the radial direction are chosen as
$r_i=50.0\tan\left[(i+0.5){0.5\pi \over 512+1} \right]$kpc.

As a first application, the RAVE solar neighbourhood escape speed
$544^{+64}_{-46}$~km/s is well-reproduced by our fully numerical
model galaxy (Fig.~1 for a typical external field of $0.01a_0$), as
analytically anticipated in Famaey et al.~(2007b). When the direction of the
external gravity changes, the escape speed changes in a narrow range
$[545,558] \, {\rm km} \, {\rm s}^{-1}$ in the solar neighborhood. We moreover point out that satellites of Milky Way-like galaxies will move on nearly Keplerian orbits at very large radii ($>100$-$500$~kpc depending on the external field strength), with a speed much below the asymptotically flat naive MOND prediction.

\begin{figure}
\resizebox{8cm}{!}{\includegraphics{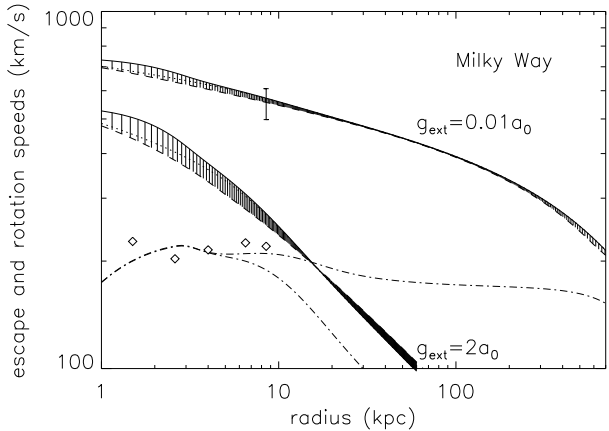}} \vskip 0.5cm
\caption{Model of a Milky Way-like galaxy in weak ($0.01a_0$) and
strong ($2a_0$) external fields.  Escape speeds in the disk plane
for various field directions (solid and dotted) are compared with
the error bar for the local escape speed measured from the RAVE
survey(Smith et. al. 2007). The predicted circular speed
curves (dot-dashed lines) are also compared with data
(diamonds)(Caldwell \& Ostriker 1981). Note that for an external field of $0.01a_0$, the transition radius between MOND and Keplerian regime is at $\sim500$~kpc, while it is at $\sim150$~kpc for an external field of $0.03a_0$.}\label{figmw}
\end{figure}

We also model NGC~1560, a benchmark Low Surface Brightness (LSB)
disk galaxy (Broeils 1992). We use an exponential stellar disk of
$1.97\times10^8\Msun$ and a multi-Gaussian gaseous component of
$1.07\times10^9\Msun$ to match the observed baryon
distribution(Broeils 1992).  A MOND Poisson solver (Ciotti,
Londrillo, \& Nipoti 2006) with $256 \times 64 \times 64$ grid
points is applied, the radial grid being $r_i=10.0
\tan\left[(i+0.5){0.5\pi \over 256+1} \right]$kpc. Again it is found
that an acceleration of $0.01a_0$ is compatible with the rotation
curve of NGC1560, which is an isolated galaxy.

\begin{figure}
\resizebox{8cm}{!}{\includegraphics{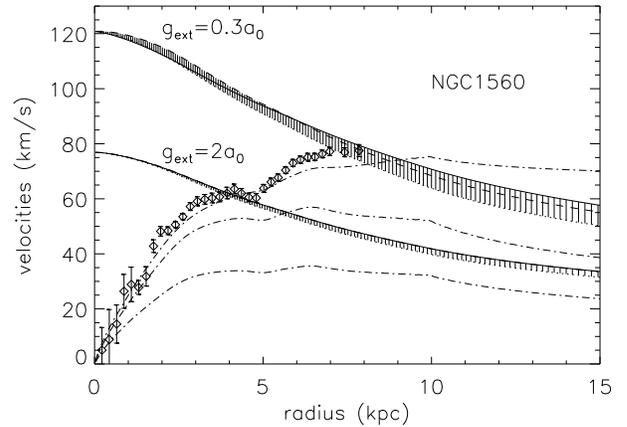}} \vskip 0.5cm
\caption{Similar to Fig.1, but for models of a NGC~1560-like LSB
galaxy. Circular speed curves for $g_{\rm ext}=0$ (no escape),
$g_{\rm ext}=0.3a_0$ (middle) and $g_{\rm ext}=2a_0$ (bottom) are
compared with the observed rotation curve of
NGC~1560.}\label{figlsb}
\end{figure}

\section{Fast-accelerating galaxies in clusters}

Now consider boosting the Milky Way's systematic acceleration {\it
suddenly} to match the environment in a galaxy cluster.
Fig.~\ref{figmw} shows for an external field of $2a_0$, the escape
speed of stars is much reduced, falling Keplerian-like $300
\sqrt{5\kpc/r}\kms$ outside 5 kpc, where half of the stars and gas
of the Milky Way are located. All dwarf satellites of the Milky Way, and
outer disk rotating with 200km/s would then barely be kept from flying away.

In fact, the instantaneous hypothetical circular speed must also be
lowered by the sudden boost of acceleration, and the outer galaxy ($>5$~kpc)
should exhibit a Keplerian falling rotation curve (Fig.1).  Outer
disk stars and gas should enter elliptical or parabolic orbits of
the same angular momentum if allowed to respond to a suddenly
reduced gravity, and precess with a preferred direction of
instantaneous systematic acceleration which thickens the disk.  In
any case, observing asymptotically flat rotation curves for purely
baryonic Milky Way-like galaxies residing in such environments would
falsify MOND. Galaxy of same luminosity would have lower velocities,
consistent with the observed trend with cluster Tully-Fisher
relation(Sanders \& McGaugh 2002).

Now let us consider suddenly boosting the acceleration of our benchmark LSB
galaxy to $\dot{v}_{com}=0.3a_0$ or $\dot{v}_{com}=2a_0$ (typical of
outer and inner parts of galaxy clusters. The real orbit of a member
galaxy would pass both regions at apocenter and pericenter
respectively). The circular speed (Fig.~\ref{figlsb} lowest dot-dashed curve) of a
fast-accelerating LSB is very much reduced. All previous
disrupting effects are even more severe on an LSB galaxy with the
escape speed falling as low as $50$~km/s.  Outer stars with original
circular speed $~50-80$ km/s would enter parabolic orbits, and inner
stars move outwards on severely elongated non-planar orbits.
Actually, the dynamics resembles a purely Newtonian disk without a
round stablizing dark halo, meaning that the galaxy would become
extremely bar-unstable (Mihos, McGaugh \& de Blok 1997). Such an LSB
would lose its MOND support and would be subject to strong
distortions, even before the traditional tidal effect becomes
important.

\section{Conclusion and discussion}

The external field effect is a generic prediction of the modified
gravity theories where the modification is acceleration-based and
violates the Strong Equivalence Principle, e.g., the relativistic
versions of the AQUAL theory of Bekenstein \& Milgrom (1984). The
effect is helpful (i) to allow high velocity stars to escape from Milky
Way-like field galaxies, and (ii) to decrease the orbital velocities of their satellite galaxies at very large radii (100 - 500 kpc), contrary to the naive expectation of MOND without external field effect, i.e. that rotation curves should be asymptotically flat. In this respect, the data of Klypin \& Prada (2007) will be very useful in the future to verify/falsify/quantify this effect thanks to detailed numerical modelling.

On the other hand the internal dynamical
structure of a field galaxy would transform when entering a cluster.
Classical relations of field galaxies, such as the Tully-Fisher relation, the galaxy luminosity functions, the Hubble type
distribution etc. are expected to modify strongly in clusters.  The
effects are most destructive for classical LSB galaxies; curiously
their field counterparts have been a legendary success for MOND in
terms of well-fitted rotation curves.

We thus argue that it would be extremely valuable to analyse the
kinematics of a {\it sample} of HSB galaxies and search for LSB
galaxies in nearby clusters using deep HI surveys. {\relax The study
of a sample of galaxies would be needed because of the uncertainty
of the determination of the real distance (as opposed to projected
distance) of a galaxy from the cluster center}. An obvious
difficulty will be that cluster galaxies are HI-deficient (Solanes
et.al. 2001). An example of such an HI database is the VIVA survey
(VLA Imaging of Virgo in Atomic Gas (Chung, van Gorkom et. al. 2007). We also predict that a future detection of any undistorted HSB
late-type disk galaxy near the center of a galaxy cluster would be
extremely surprising in the context of MOND.  A null-detection of
thin LSB disks is predicted in clusters because they most probably
have been turned into gas-poor dwarf ellipticals if not fully
disrupted.  Low surface density gas in galaxies also suffers ram
pressure stripping while moving in gaseous clusters; gas is easily
stripped in the reduced MONDian internal gravity, further reducing
available mass for self-gravity.

Surely similar effects occur in the context of cored Dark Halos.
Some simulations show that LSB disks and dwarf irregulars get
harassed (Moore 1999) and transformed into dwarf ellipticals or
ultra-compact dwarf ellipticals (Evstigneeva et al. 2007; Cortese et al. 2007) in the densest part of the cluster coinciding with the
region where the external field is the highest. To our knowledge the
properties of cluster disk galaxies (such as their Tully-Fisher
relation) have not been extensively simulated. The important
assumption of existing simulations is a large core for the CDM halo
of the LSB; the harassment becomes much less effective if the
cluster member LSB started with a dense CDM cusp density (Lucio
Mayer 2007, private communications).

As for the gas-poor non-rotating dwarf spheroidals (e.g., Sextans),
they are expected to have a central CDM density of $\sim
0.1\msun\pc^{-3}$, a factor of 100 denser than the galaxy cluster,
hence might survive the tidal harassment in CDM.  If in MOND a
spheroidal object of $M=2 \times 10^5\msun$, and half mass radius of $s
\sim 500$pc is suddenly introduced into a galaxy cluster, it would
have a central binding energy of only $\sim {G M \over
s}(1+\Delta)^{-2/3}\mu_m^{-1} \sim (5\kms)^2$, much less than its
initial internal random motion energy ${3 \times (10\kms)^2 \over
2}$, hence is quickly dispersed (perhaps anisotropically).
\footnote{For similar reasons open clusters in the solar
neighbourhood are predicted internally unbound in MOND.} In short, any discovery
of a sample of classical LSB galaxies in clusters would favor cuspy
CDM, and falsify MOND or cored Dark Halos.

The tidal harassment effect exists in MOND as well (Zhao 2005); cluster
galaxies suffer from tides in addition to the unique destructive
effect of the external field. An even more curious distortion to the
MONDian LSB or HSB disk happens when the disk is mis-aligned by an
angle $\theta_m$ with the instantaneous direction of the external
field, which generally changes amplitude and direction along the
orbit of an LSB on time scales of 0.2-1Gyr. The elliptical potential
of Eq.(4) creates a differential force with a component normal to
the disk, hence a specific torque $-{\bf r} \times \grad \Phi$. This
causes differential precession of the disk angular momentum vector
with an angular speed proportional to $\mu_m^{-1} \sqrt{G M/r^3}
\Delta \sin(2\theta_m)$; an LSB disk is likely shredded by one
precession and a HSB disk is thickened. The {\it precession,
asymmetric dilation and reduction of {\it inner} circular velocity
curves} (cf. Fig.2) are confirmed by N-body simulations using the
code of the Paris group (Tiret \& Combes 2007) in MOND, but are
generally forbidden by Newtonian laws in the Dark Matter framework.

\begin{acknowledgments}
We thank Luca Ciotti, Pasquale Londrillo, Carlo Nipoti for
generously sharing their code, and Tom Ritchler for helpful comments
on an earlier manuscript.
\end{acknowledgments}


\begin{thebibliography}{}
\bibitem{AM07} Angus G.W., McGaugh S.S., 2007, arXiv:0704.0381
\bibitem{A07} Angus G.W., Shan H., Zhao H.S., Famaey B., 2007, ApJ, 654, L13
\bibitem{Bek04} Bekenstein J.D., 2004, Phys. Rev. D, 70, 083509
\bibitem{B07} Bekenstein J.D., 2006, Contemporary Phys., 47, 387
\bibitem{BM84} Bekenstein J.D., Milgrom M., 1984, ApJ, 286, 7
\bibitem{B92} Broeils A.H., 1992, A\&A, 256, 19
\bibitem{CO81} Caldwell J.A.R., Ostriker J.P., 1981, ApJ, 251, 61
\bibitem{CZ} Chen D.M., Zhao H.S., 2006, ApJ, 650, L9
\bibitem{Chung07} Chung A., et al., 2007, ApJ, 659, 115
\bibitem{C06} Ciotti L., Londrillo P., Nipoti C., 2006, ApJ, 640, 741
\bibitem{Cortese} Cortese L., et al., 2007, MNRAS, 376, 157
\bibitem{Evst} Evstigneeva E.A., et al., 2007, AJ, 133, 1722
\bibitem{FB05} Famaey B., Binney J., 2005, MNRAS, 363, 603
\bibitem{F07} Famaey B., Gentile G., Bruneton J.P., Zhao H.S., 2007a, Phys. Rev. D75, 063002
\bibitem{FBZ07} Famaey B., Bruneton J.P., Zhao H.S., 2007b, MNRAS, 377, L79 
\bibitem{Fring} Famaey B., Angus G.W., Gentile G., Zhao H.S., 2007c, arXiv:0706.1279
\bibitem{G07} Gentile G., Salucci P., Klein U., Granato G. L., 2007a, MNRAS, 375, 199
\bibitem{GF07} Gentile G., Famaey B., Combes F., Kroupa P., Zhao H.S., Tiret O., 2007b, arXiv:0706.1976
\bibitem{KP07} Klypin A., Prada F., 2007, arXiv:0706.3554
\bibitem{McGa} McGaugh S., et al., 2007, ApJ, 659, 149
\bibitem{MMB} Mihos J.C., McGaugh S.S., de Blok W.J.G., 1997, ApJ, 477, L79 
\bibitem{M83} Milgrom M., 1983, ApJ, 270, 365
\bibitem{M86} Milgrom M., 1986, ApJ, 302, 617
\bibitem{M02} Milgrom M., 2002, ApJ, 577, 75
\bibitem{MS03} Milgrom M., Sanders R.H., 2003, ApJ, 599, 25
\bibitem{M99} Moore B., 1999, Roy Soc. of London Phil. Tr. A., 357, 1763
\bibitem{N07} Nipoti C., Londrillo P., Zhao H.S., Ciotti L., 2007, MNRAS in press, arXiv:0704.0740
\bibitem{P05} Pointecouteau E., Arnaud M., Pratt G.W., 2005, A\&A, 435, 1
\bibitem{R03} Robin A.C., Reyle C., Derriere S., Picaud S., 2003, A\&A, 409, 523
\bibitem{SM02} Sanders R.H., McGaugh S.S., 2002, ARA\&A, 40, 263
\bibitem{SN07} Sanders R.H., Noordermeer E., 2007, astro-ph/0703352
\bibitem{Sk} Skordis C., et al., 2006, Phys. Rev. Lett. 96, 1301
\bibitem{RAVE} Smith M.C., et. al., 2007, astro-ph/0611671
\bibitem{Sol01} Solanes J.M. et.al., 2001, ApJ, 548, 97
\bibitem{Tiret07} Tiret O., Combes F., 2007, A\&A, 464, 517
\bibitem{Z07} Zhao H.S., 2007, astro-ph/0610056
\bibitem{ZB06} Zhao H.S., Bacon D., et al., 2006, MNRAS, 368, 171
\bibitem{ZF06} Zhao H.S., Famaey B., 2006, ApJ, 638, L9
\bibitem{ZT05} Zhao H.S., Tian L., 2006, A\&A, 450, 1005
\bibitem{ZlosnikFS07} Zlosnik T., Ferreira P., Starkman G., 2007, Phys. Rev. D75, 4017
\end{thebibliography}
\end{document}